\documentclass[aps,prb,reprint]{revtex4-1} 

\usepackage{amsmath}  
\usepackage{amsfonts} 
\usepackage{graphicx} 
\usepackage{comment}
\usepackage{tikz}
\usetikzlibrary{arrows.meta, positioning}
\usepackage{xr-hyper} \usepackage[colorlinks=true, linkcolor=blue, citecolor=blue, urlcolor=blue]{hyperref}
\begin{document}
\title{From Wave Scattering to Bloch Bands: \\ A Time-Domain Approach to Band Formation in Periodic Media}

\author{Nishant Kashyap}
\author{Amit Tanwar}
\author{Pragati Ashdhir}
\email{pragatiashdhir@hinducollege.ac.in}
\affiliation{
Department of Physics, 
Hindu College, University of Delhi, Delhi-110007, India
}
\author{Vivek T. Ramamoorthy}
\affiliation{
School of Physics, Engineering, and Computer Science, 
University of Hertfordshire, Hatfield AL10 9AB, United Kingdom
}
\begin{abstract}
Band formation in periodic media is a central topic in undergraduate solid-state physics, typically introduced through Bloch’s theorem as an eigenvalue problem in reciprocal space for infinitely periodic systems. While mathematically elegant, this formulation can appear abstract: it assumes an idealized infinite lattice, shifts attention away from real-space wave dynamics, and presents band structures as static results rather than emergent consequences of wave propagation. Consequently, students often struggle to relate band gaps to familiar physical phenomena such as reflection, transmission, and interference, leading to a disconnect between formal band theory and observable wave behavior. We present a computational framework that addresses this gap by reconstructing band formation directly from time-domain wave propagation in finite periodic systems. Using a staggered-grid finite-difference time-domain scheme for elastic waves, a broadband excitation is propagated through a layered medium to obtain its transmission spectrum. From this, students extract the Bloch dispersion relation and observe spatial attenuation in band-gap regions, revealing the roles of multiple scattering and phase coherence. This approach provides a physically transparent pathway to band theory and enables exploration of finite-size effects, disorder, and defect-localized modes within a unified computational framework. Implemented through compact code and guided exercises, the method offers an accessible and versatile pedagogical tool, while also equipping students with transferable skills in numerical modeling of wave phenomena across disciplines.

\end{abstract}

\maketitle 
\section{Introduction}
\label{sec:introduction}

Wave phenomena appear across many areas of physics, from mechanical vibrations and elastic waves to electromagnetic radiation and quantum matter waves. A particularly important manifestation arises in periodic structures, where wave propagation organizes into allowed bands separated by forbidden regions. In undergraduate courses, this behavior is most often introduced through models such as the Kronig--Penney potential\cite{kronig1931} and Bloch’s theorem,\cite{bloch1929} which treat band gaps as spectral properties of an ideal infinite lattice.\cite{kittel2004,ashcroft1976} Although analytically appealing, this presentation can make it difficult for students to connect the abstract spectral picture with the time-domain wave dynamics they already understand.

In this article we present a complementary computational perspective that emphasizes the dynamical origin of band formation. Using a compact finite-difference time-domain (FDTD)\cite{taflove2005} scheme written in velocity--stress variables, one can follow how repeated scattering in a layered medium builds the collective propagation behavior associated with periodic systems. A broadband excitation reveals the system’s transmission and attenuation characteristics, showing how multiple scattering progressively organizes wave propagation. This viewpoint complements the usual algebraic treatment of periodic media by grounding dispersion relations in causally transparent wave dynamics, while remaining simple enough to be implemented by upper-division undergraduate students.

To illustrate this approach in a concrete physical setting, we consider longitudinal elastic waves in periodic media. Structures designed to control the propagation of such waves are known as phononic crystals.\cite{kushwaha1993} These engineered materials exhibit band gaps for acoustic and elastic waves due to periodic variations in density or elastic stiffness, closely paralleling the electronic band structure produced by periodic potentials in solids.\cite{kittel2004,ashcroft1976} This framework is pedagogically convenient because the fundamental fields---particle velocity and stress---correspond directly to familiar mechanical quantities. At the same time, the underlying physical ideas extend beyond elastic systems: the same repeated-scattering mechanism governs wave propagation in a wide range of periodic media, including electromagnetic waves in photonic structures\cite{mcgurn2022} and quantum particles in crystalline solids. Phononic crystals are also of growing technological interest, with applications in vibration isolation, acoustic filtering, and wave guiding.\cite{liu2020}

The central aim of this paper is to present a compact computational framework through which band formation in periodic media is reconstructed directly from time-domain multiple scattering in finite layered systems. In this approach, the emergence of band structure, Bloch dispersion, and spatial attenuation can be traced explicitly to underlying wave dynamics. The instructional strategy is to begin with simple scattering processes and progressively build toward the collective behavior of periodic systems. The approach combines conceptual understanding of wave physics with practical experience in numerical modeling through a minimal, accessible, and transferable code structure. Selected sections include short computational exercises that encourage a learning-by-doing approach through direct simulation. Fully documented Python codes are provided to ensure reproducibility and to support implementation by advanced undergraduate students.

\section{From Displacement to Velocity--Stress Formulation}
\label{sec:displacement_to_fields}
Mechanical waves in solids are commonly introduced in terms of the displacement field $u(x,t)$. In a homogeneous elastic medium, the motion obeys the familiar one-dimensional wave equation\cite{achenbach2012}
\begin{equation}
\label{eq:wave_uniform}
\frac{\partial^2 u}{\partial t^2} = c^2 \frac{\partial^2 u}{\partial x^2}, 
\qquad c^2 = \frac{E}{\rho},
\end{equation}
where $E$ is the elastic modulus and $\rho$ the mass density. This form assumes a uniform medium in which the propagation speed remains constant.

Many physical systems of interest depart from this idealization. When the material properties vary along the direction of propagation, the governing equation takes the more general form\cite{achenbach2012}
\begin{equation}
\label{eq:wave_variable}
\rho(x)\frac{\partial^2 u}{\partial t^2}
=
\frac{\partial}{\partial x}\!\left[E(x)\frac{\partial u}{\partial x}\right],
\end{equation}
which describes wave propagation in spatially structured media. In the phononic crystals considered here, the density $\rho(x)$ and elastic modulus $E(x)$ vary in a piecewise periodic manner along the propagation direction, reflecting the alternating sequence of constituent layers. These spatial variations modify the local wave speed and produce reflections at material interfaces, whose cumulative effect governs the propagation behavior examined later in this work.

While compact, the displacement formulation of Eq.~\eqref{eq:wave_variable} does not explicitly reveal the dynamical mechanism underlying wave propagation. A clearer physical picture emerges when the motion is expressed in terms of fields representing momentum transport and restoring forces. To make this structure explicit, it is convenient to recast the second-order equation into an equivalent first-order system. Introducing the particle velocity $v=\partial u/\partial t$ and the stress $\sigma=E(x)\,\partial u/\partial x$, the elastodynamic equations can be written as the coupled first-order system\cite{landau1986,achenbach2012}
\begin{subequations}
\label{eq:vel_stress_form}
\begin{align}
\frac{\partial v}{\partial t} &= \frac{1}{\rho(x)}\frac{\partial \sigma}{\partial x}, \label{eq:vel_eq}\\
\frac{\partial \sigma}{\partial t} &= E(x)\frac{\partial v}{\partial x}. \label{eq:stress_eq}
\end{align}
\end{subequations}

In this representation, wave propagation appears as the mutual evolution of two interdependent fields. Spatial variations of stress accelerate material elements through Eq.~\eqref{eq:vel_eq}, while spatial variations of velocity modify the local stress through Eq.~\eqref{eq:stress_eq}. Wave motion thus emerges from a continual local exchange between kinetic and elastic energy. The material parameters $E$ and $\rho$ govern the strength of this coupling and therefore determine how efficiently energy is transmitted through the medium.

This dynamical structure closely parallels that of Maxwell’s equations for electromagnetic waves, where electric and magnetic fields similarly drive one another through first-order spatial derivatives. The analogy highlights a general feature of wave physics: propagation arises from the coupled evolution of complementary fields. In elastodynamics, however, the variables retain a direct mechanical interpretation: the particle velocity describes the motion of material elements, while the stress represents the internal forces transmitted through the medium.

 The first-order elastodynamic system in Eq.~\eqref{eq:vel_stress_form} provides the natural starting point for the numerical formulation developed in this work. In the following section, we construct a staggered-grid finite-difference scheme that respects this coupled structure and enables efficient simulation of wave propagation in media with spatially varying material properties.

\section{A Staggered-Grid Formulation for Elastodynamic Waves}
\label{sec:stag_grid_formulation}

In its most familiar form, the finite-difference method is introduced through Taylor-series expansions, leading naturally to a collocated grid in which all field variables are defined at the same spatial nodes and time levels.\cite{leveque2007} For problems governed by a single differential equation in a homogeneous medium, this arrangement is often sufficient. However, difficulties arise in structured media where material parameters vary discontinuously, particularly for coupled first-order systems such as the velocity--stress formulation introduced in Eq.~\eqref{eq:vel_stress_form}.

In such systems, each field evolves through spatial variations of the other: stress gradients drive changes in velocity, while velocity gradients determine the evolution of stress. When both variables are stored at identical grid locations, discrete derivatives may span material interfaces where coefficients change abruptly, leading to an inaccurate representation of momentum and stress balance.\cite{taflove2005,patankar1980} A staggered grid avoids this difficulty by storing the variables at interleaved positions. Velocities are naturally associated with the motion of material elements, while stresses represent forces transmitted between neighboring elements. Placing these quantities at offset spatial locations therefore aligns the discrete variables with their physical roles and improves the representation of interfaces in structured media.

\begin{figure}
\centering
\includegraphics[width=\linewidth]{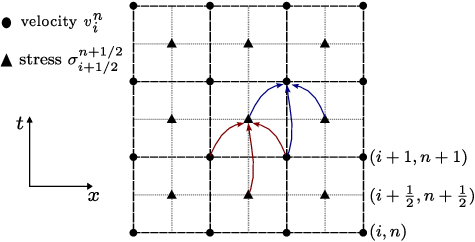}
\caption{Schematic of the staggered-grid arrangement for the one-dimensional velocity--stress formulation. Velocity is defined at integer spatial and temporal nodes, while stress is defined at half-integer locations. The arrows illustrate the interleaved leapfrog update of the two fields.}
\label{fig:1d_staggered_grid}
\end{figure}

The staggered-grid finite-difference scheme provides a natural framework for this purpose. Originally introduced by Yee\cite{yee1966} in the context of electromagnetic wave propagation, this approach defines coupled variables at interlaced positions in space and time. For the one-dimensional elastodynamic system considered here, velocity and stress are defined at distinct spatial locations and time levels, reflecting their causal coupling. Specifically, the velocity $v$ is defined at integer spatial nodes and time steps $(i,n)$,
\begin{equation}
v_i^n \approx v(i\Delta x, n\Delta t),
\end{equation}
while the stress $\sigma$ is defined at half-integer spatial nodes and time levels $(i+\tfrac{1}{2}, n+\tfrac{1}{2})$,
\begin{equation}
\sigma_{i+1/2}^{\,n+1/2} \approx \sigma\!\left((i+\tfrac{1}{2})\Delta x,(n+\tfrac{1}{2})\Delta t\right),
\end{equation}
where $\Delta x$ and $\Delta t$ denote the spatial and temporal step sizes. The spatial and temporal staggering of the two fields is illustrated schematically in Fig.~\ref{fig:1d_staggered_grid}, making explicit that each field is updated using information from adjacent locations of the other.

With this arrangement, temporal updates occur in an alternating (leapfrog) manner.\cite{taflove2005} The time derivatives are approximated as
\begin{equation}
\left.\frac{\partial v}{\partial t}\right|_{i}^{\,n+1/2}
\approx
\frac{v_i^{\,n+1}-v_i^{\,n}}{\Delta t},
\end{equation}
and
\begin{equation}
\left.\frac{\partial \sigma}{\partial t}\right|_{i+1/2}^{\,n}
\approx
\frac{\sigma_{i+1/2}^{\,n+1/2}-\sigma_{i+1/2}^{\,n-1/2}}{\Delta t}.
\end{equation}

Similarly, the spatial derivatives are approximated by
\begin{equation}
\left.\frac{\partial v}{\partial x}\right|_{i+1/2}^{\,n}
\approx
\frac{v_{i+1}^{\,n}-v_i^{\,n}}{\Delta x},
\end{equation}
and
\begin{equation}
\left.\frac{\partial \sigma}{\partial x}\right|_{i}^{\,n+1/2}
\approx
\frac{\sigma_{i+1/2}^{\,n+1/2}-\sigma_{i-1/2}^{\,n+1/2}}{\Delta x}.
\end{equation}
Each derivative is evaluated at the location where the corresponding update is defined, ensuring that the discrete system respects the structure of the continuous equations.

Substituting these approximations into Eq.~\eqref{eq:vel_stress_form}, the stress field advances from $n-1/2$ to $n+1/2$ using velocity values at time $n$,
\begin{equation}
\label{eq:1DStress_update}
\sigma_{i+1/2}^{\,n+1/2}
=
\sigma_{i+1/2}^{\,n-1/2}
+
\left(\frac{E_{i+1/2}\Delta t}{\Delta x}\right)
\left(v_{i+1}^{\,n}-v_i^{\,n}\right),
\end{equation}
and the velocity field subsequently advances from $n$ to $n+1$ using the updated stress values,
\begin{equation}
\label{eq:1DVelocity_update}
v_i^{\,n+1}
=
v_i^{\,n}
+
\left(\frac{\Delta t}{\rho_i\Delta x}\right)
\left(\sigma_{i+1/2}^{\,n+1/2}-\sigma_{i-1/2}^{\,n+1/2}\right).
\end{equation}

The alternating updates between the two fields correspond directly to the interleaved sequence indicated in Fig.~\ref{fig:1d_staggered_grid}, where, stress is first updated from neighboring velocities, and the resulting stresses then drive the subsequent velocity update.

Here $E_{i+1/2}$ denotes the elastic modulus at the interface between nodes $i$ and $i+1$, evaluated using the harmonic average of adjacent cell-centered values to ensure consistent stress continuity across material interfaces.\cite{patankar1980} The density $\rho_i$ is evaluated at the integer node $i$, consistent with the location of the velocity field. Equations~\eqref{eq:1DStress_update} and \eqref{eq:1DVelocity_update} constitute the discrete staggered-grid representation of the coupled system in Eq.~\eqref{eq:vel_stress_form} and form the core of the numerical solver used throughout this work.

\section{Source Excitation and Absorbing Boundaries}
\label{sec:initial_boundary}

Numerical simulations of wave propagation require a mechanism to introduce energy into the system. In the present formulation, this is achieved through a time-dependent source that generates a localized wave packet within the computational domain. Transient excitations are particularly useful because their spectral content can be controlled directly, allowing the frequency bandwidth to be chosen consistently with the numerical resolution and stability constraints discussed in Sec.~\ref{sec:numerical_stability}.

A convenient choice for a localized transient excitation is the Ricker wavelet,\cite{schneider2010} which has zero mean and a well-defined central frequency. The zero-mean property avoids residual static displacement after the pulse has passed, while the dominant frequency can be tuned to probe specific spectral regions. Its temporal form is
\begin{equation}
s(t) = A \left[1 - 2\pi^2 f_0^2 (t - t_0)^2 \right]
\exp\!\left[-\pi^2 f_0^2 (t - t_0)^2\right],
\end{equation}
where $A$ is the amplitude, $f_0$ the central frequency, and $t_0$ a temporal offset controlling the location of the pulse.

When a broader and approximately uniform spectral content is required, a windowed sinc pulse is more appropriate.\cite{oppenheim1999} A band-limited excitation with cutoff frequency $f_{\text{cut}}$ may be written as
\begin{equation}
s(t) = A\,\mathrm{sinc}\!\left[2 f_{\text{cut}}(t - t_0)\right]\, W(t),
\end{equation}
where $\mathrm{sinc}(x)=\sin(\pi x)/(\pi x)$ and $W(t)$ is a Blackman window of duration $2t_0$,
\begin{equation}
W(t)=
\begin{cases}
\begin{aligned}
0.42 
&- 0.5 \cos\!\left(\dfrac{\pi (t - t_0)}{t_0}\right) \\
\quad &+ 0.08 \cos\!\left(\dfrac{2\pi (t - t_0)}{t_0}\right),
\end{aligned}
& |t - t_0| \le t_0, \\[8pt]
0, & \text{otherwise}.
\end{cases}
\end{equation}

\begin{figure}
\centering
\includegraphics[width=\linewidth]{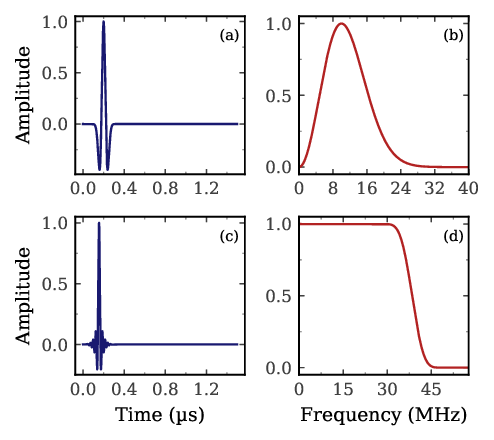}
\caption{Comparison of narrowband and broadband source excitations. (a) Ricker wavelet in the time domain. (b) Corresponding spectrum centered at $f_0=10\,\mathrm{MHz}$. (c) Blackman-windowed sinc pulse. (d) Approximately uniform spectrum extending up to the prescribed cutoff frequency $f_{\text{cut}}\approx38.5\,\mathrm{MHz}$.}
\label{fig:sources}
\end{figure}

The sinc factor produces an approximately flat spectrum up to $f_{\text{cut}}$, while the Blackman window confines the pulse in time and suppresses spectral side lobes. In practice, the choice of excitation is guided by the desired spectral content: the Ricker wavelet provides a smooth narrowband pulse centered at $f_0$, whereas the windowed sinc pulse enables broadband excitation, allowing transmission spectra to be obtained over a wide frequency range in a single simulation. This comparison highlights how the choice of source directly determines the spectral information accessible in a single simulation. The time-domain signals and their corresponding spectra are compared in Fig.~\ref{fig:sources}.

In addition to source excitation, numerical simulations must address the treatment of the computational boundaries. The physical configuration considered here corresponds to a structured region embedded within an effectively unbounded host medium, with waves incident from one side and transmitted to the other. Ideally, the surrounding medium would extend indefinitely so that only scattering from the structured region influences the observed response. In practice, however, the computational domain must be finite, and artificial reflections from its edges must be suppressed.

This is achieved by applying absorbing boundary conditions (ABCs) at the edges of the grid, so that outgoing waves leave the computational domain with minimal reflection. Two commonly used approaches are Mur’s absorbing boundary condition\cite{mur1981} and the perfectly matched layer (PML).\cite{berenger1994} Mur’s first-order ABC is particularly simple to implement, as it requires only local update relations at the boundary nodes. In one-dimensional problems, where waves reach the boundaries at normal incidence and the surrounding medium is homogeneous, this condition is often sufficient to approximate an open domain.

This absorbing boundary condition is derived from the one-way wave equation, which admits only outgoing waves at the boundary. At the right boundary, only right-traveling waves are permitted, while at the left boundary only left-traveling waves are allowed. Discretizing this condition using first-order finite differences yields the update relation for the boundary velocity node
\begin{equation}
v_{N}^{\,n+1} =
v_{N-1}^{\,n} +
\frac{c\Delta t - \Delta x}{c\Delta t + \Delta x}
\left( v_{N-1}^{\,n+1} - v_{N}^{\,n} \right),
\end{equation}
where $N$ denotes the rightmost grid point. The corresponding condition at the left boundary ($i=0$) becomes
\begin{equation}
v_{0}^{\,n+1} =
v_{1}^{\,n} +
\frac{c\Delta t - \Delta x}{c\Delta t + \Delta x}
\left( v_{1}^{\,n+1} - v_{0}^{\,n} \right).
\end{equation}

Here $c=\sqrt{E/\rho}$ denotes the wave speed of the host medium adjacent to the boundary. These relations are applied after the interior velocity update at each time step, allowing waves reaching the boundaries to exit the domain with minimal reflection. A detailed derivation of these discrete boundary relations is provided in Sec.~\ref{supsec:ABCs} of the Supplementary Notes,\cite{SuppMat} where the one-way wave equation and its finite-difference approximation are developed explicitly.

The effectiveness of Mur’s condition depends on the discretization and is strictly valid only for normally incident waves in a homogeneous medium.\cite{hwang2007} In more complex situations such as oblique incidence, broadband excitation, or multidimensional propagation, numerical reflections and instabilities may accumulate and more robust absorbing layers are preferred. The perfectly matched layer (PML) is widely used because it can absorb waves over a broad frequency range and for arbitrary angles of incidence. The derivation and implementation of the PML formulation are presented in Sec.~\ref{supsec:ABCs} of the Supplementary Notes.\cite{SuppMat}

These elements—controlled source excitation and minimally reflective boundaries—ensure that the simulated wave dynamics faithfully represent propagation in an effectively infinite medium.

\section{Resolution, Stability, and Numerical Dispersion}
\label{sec:numerical_stability}

The staggered-grid scheme is second-order accurate in both space and time, $\mathcal{O}(\Delta x^2,\Delta t^2)$ (see Sec.~\ref{supsec:fdtd_accuracy} of the Supplementary Notes \cite{SuppMat}). In practice, however, the numerical parameters $\Delta x$ and $\Delta t$ must be chosen carefully to realize this formal accuracy. The spatial grid must resolve the shortest wavelength present in the simulation, while the time step must satisfy the stability constraint of the explicit scheme.

The smallest wavelength to be resolved is
\begin{equation}
\lambda_{\min} = \frac{c_{\min}}{f_{\max}},
\end{equation}
where $c_{\min}$ is the smallest wave speed in the structure and $f_{\max}$ is the highest significant frequency component of the excitation. For a Ricker wavelet, $f_{\max}$ is related to the chosen central frequency (typically $f_0 \approx 0.30\,f_{\max}$), whereas for the windowed sinc pulse it is set directly by the cutoff frequency (typically $f_{\text{cut}} \approx 0.75\,f_{\max}$). The spatial grid spacing is therefore selected through a points-per-wavelength (PPW) requirement,
\begin{equation}
N_\lambda = \frac{\lambda_{\min}}{\Delta x}.
\end{equation}
In practice, $N_\lambda \gtrsim 10$--15 is typically required for acceptable phase accuracy,\cite{pennec2018}
while strongly heterogeneous systems often require $20$--$60$ PPW.\cite{wei2015,schroder2002}

Once $\Delta x$ is fixed, the time step must satisfy the Courant--Friedrichs--Lewy (CFL) stability condition,\cite{schneider2010}
\begin{equation}
\Delta t \le \frac{\Delta x}{c_{\max}},
\end{equation}
where $c_{\max}$ is the largest wave speed in the computational domain. Physically, this condition ensures that the numerical wave cannot travel farther than one grid cell during a single time step, so that information propagates across the grid in a stable and causal manner ($c_{\max}\Delta t \le \Delta x$). Thus $c_{\min}$ determines the spatial resolution, while $c_{\max}$ constrains the stability of the time integration.

Stability alone does not eliminate phase error. In a discrete simulation, waves propagate with a numerical phase velocity $v_{\mathrm{num}}$, which may differ from the physical wave speed $c$. This discrepancy is known as numerical dispersion.\cite{taflove2005} For the one-dimensional second-order staggered-grid scheme, the corresponding discrete dispersion relation is
\begin{equation}
\frac{1}{c\Delta t}\sin\!\left(\frac{\omega\Delta t}{2}\right)
=
\frac{1}{\Delta x}\sin\!\left(\frac{k\Delta x}{2}\right),
\end{equation}
where $\omega$ is the angular frequency and $k$ is the wavenumber of a plane-wave solution $e^{i(kx-\omega t)}$. Solving this relation gives the normalized numerical phase velocity
\begin{equation}
\frac{v_{\mathrm{num}}}{c}
=
\frac{\omega}{ck}
=
\frac{2}{\zeta k\Delta x}
\arcsin\!\left[
\zeta \sin\!\left(\frac{k\Delta x}{2}\right)
\right],
\end{equation}
where $\zeta = c\Delta t/\Delta x$ is the CFL number and $k\Delta x = 2\pi/N_\lambda$ relates the wavenumber to the spatial resolution.

\begin{figure}
\centering
\includegraphics[width=\linewidth]{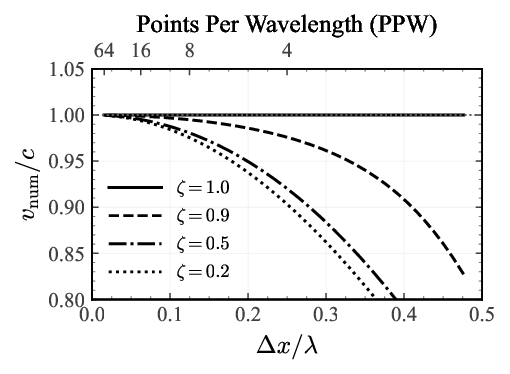}
\caption{Normalized numerical phase velocity $v_{\mathrm{num}}/c$ for the second-order staggered grid as a function of $\Delta x/\lambda = 1/N_\lambda$, shown for several CFL numbers $\zeta = c\Delta t/\Delta x$. Phase error decreases with increasing spatial resolution and as $\zeta$ approaches unity.}
\label{fig:dispersion_ppw}
\end{figure}

The implications of this relation are illustrated in Fig.~\ref{fig:dispersion_ppw}. The figure shows the ratio $v_{\mathrm{num}}/c$ as a function of $\Delta x/\lambda = 1/N_\lambda$ for several values of $\zeta$. As the spatial resolution increases ($N_\lambda \rightarrow \infty$), the numerical phase velocity approaches the physical wave speed. For a fixed spatial resolution, choosing $\zeta$ close to unity (that is, selecting $\Delta t$ near the CFL limit) reduces phase error.

If the spatial grid is too coarse, the numerical phase velocity deviates from $c$, leading to numerical dispersion. In time-domain simulations, this manifests as pulse broadening and the appearance of small trailing oscillations even in a homogeneous medium. This numerical dispersion is purely a discretization effect and should be distinguished from the physical dispersion that arises in periodic media, which is the central focus of this work.

In heterogeneous systems such as layered phononic crystals, different materials support waves that travel at different speeds. Since an explicit scheme must remain stable everywhere in the domain, the global time step $\Delta t$ is chosen using the largest wave speed $c_{\max}$. This introduces a local Courant number
\begin{equation}
\zeta(x) = \frac{c(x)\Delta t}{\Delta x},
\end{equation}
which varies across the structure.

In regions where the local wave speed is much smaller than $c_{\max}$, the effective CFL number satisfies $\zeta(x) \ll 1$. As shown in Fig.~\ref{fig:dispersion_ppw}, smaller values of $\zeta$ increase numerical phase error. Consequently, choosing $\Delta t$ based on $c_{\max}$ can reduce phase accuracy in slower layers. One remedy is to reduce the spatial grid spacing $\Delta x$, which increases the points-per-wavelength resolution in all layers and improves phase accuracy, albeit at increased computational cost. Alternatively, higher-order spatial discretizations can be used to reduce phase error for a given grid resolution.\cite{moczo2000}

Once the time-domain fields are recorded, spectral quantities such as transmission are obtained by Fourier transforming the signals.\cite{oppenheim1999} The achievable frequency resolution is determined by the total simulation time $T_f$,
\begin{equation}
\Delta f = \frac{1}{T_f}.
\end{equation}
Accurate identification of spectral features therefore requires sufficiently long time integration.

Finally, the numerical scheme is explicit and involves only nearest-neighbor updates. Field arrays from two time levels are recycled in memory, keeping storage requirements minimal. The update loops are therefore well suited for efficient implementation, including just-in-time compilation approaches such as \texttt{numba},\cite{lam2015} without altering the underlying formulation.

\section{Simulation Workflow and Implementation}
\label{sec:implementation}

\begin{figure}
\centering
\begin{tikzpicture}[
node distance=5mm and 8mm,
box/.style={
draw,
rounded corners,
align=center,
minimum width=3.5cm,
minimum height=6mm,
font=\small
},
loopbox/.style={
draw,
rounded corners,
align=center,
minimum width=7.2cm,
minimum height=6mm,
font=\small,
thick
},
arrow/.style={->, thick}
]

\node[box] (structure) {Define layered structure\\ $\rho(x),\,E(x)$};

\node[box, right=of structure] (freq)
{Choose frequency range\\$f_{\max}$};

\node[box, below=of freq] (grid) {Choose grid parameters\\ $\Delta x$ (PPW) \& $\Delta t$ (CFL)};

\node[box, left=of grid] (init) {Initialize fields\\ $v=0,\;\sigma=0$};

\node[loopbox, below=8mm of grid, xshift=-2.2cm] (looptitle) {Time stepping (FDTD loop)};

\node[box, below=of looptitle] (stress) {Update stress and velocity\\ using Eq.~\eqref{eq:1DStress_update} and Eq.~\eqref{eq:1DVelocity_update}};

\node[box, below=of stress] (source_abc) {Inject source and apply\\ absorbing boundary conditions};

\node[box, below=of source_abc] (record) {Record signals\\(incident \& transmitted)};

\node[box, below=of record] (fft) {Post-processing:\\ FFT $\rightarrow$ transmission spectrum};

\draw[arrow] (structure) -- (init);
\draw[arrow] (freq) -- (grid);

\draw[arrow] (grid) -- (looptitle);
\draw[arrow] (init) -- (looptitle);

\draw[arrow] (looptitle) -- (stress);
\draw[arrow] (stress) -- (source_abc);
\draw[arrow] (source_abc) -- (record);
\draw[arrow] (record) -- (fft);

\end{tikzpicture}

\caption{Workflow of the time-domain simulation. The layered structure and frequency range determine the spatial and temporal discretization through the PPW and CFL conditions. The staggered-grid FDTD loop advances the stress and velocity fields while injecting the source and applying absorbing boundaries. Recorded signals are Fourier transformed to obtain transmission spectra.}
\label{fig:fdtd_flowchart}
\end{figure}
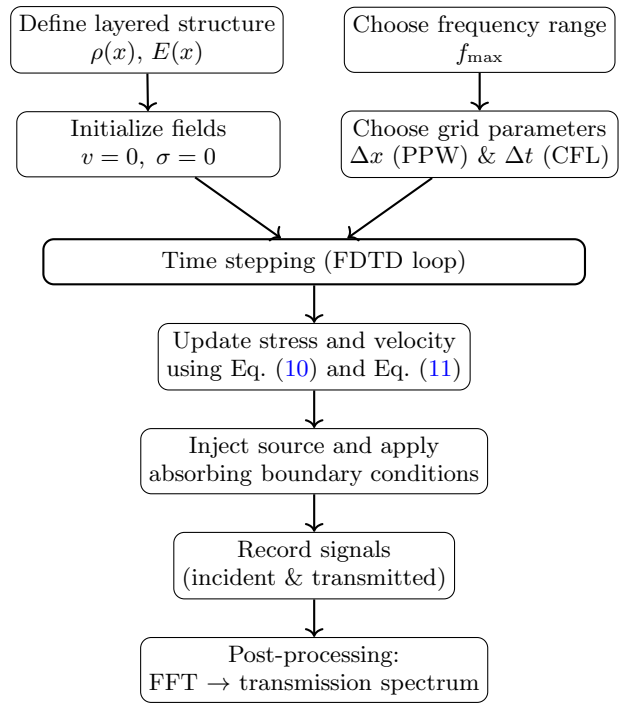

The staggered-grid discretization, source excitation, boundary treatment, and stability constraints together define a complete time-domain solver for elastic wave propagation. The overall computational procedure used in this work is summarized schematically in Fig.~\ref{fig:fdtd_flowchart}. Starting from the definition of the layered structure and the choice of numerical parameters, the algorithm advances the coupled velocity and stress fields in time while injecting the source and applying absorbing boundaries at the edges of the computational domain. This sequence of steps mirrors the logical progression followed by students when implementing the method, moving from physical modeling to numerical simulation and ultimately to spectral analysis.

This workflow highlights an important conceptual point: frequency-domain information such as transmission spectra and band structure emerges naturally from time-domain wave evolution, rather than being imposed through an eigenvalue formulation.

Because the formulation operates entirely in the time domain, spectral quantities are obtained by transforming the simulated signals to the frequency domain using Fourier analysis. In the present implementation, the recorded time-domain signals are processed using the fast Fourier transform (FFT) to obtain transmission spectra. A pedagogical discussion of the fast Fourier transform (FFT) and its applications is provided in Ref.~\onlinecite{ashdhir2021}.

Additional materials supporting the simulations are provided in the Supplementary Material,\cite{SuppMat} including animations illustrating the time-domain evolution of the fields and the complete simulation codes used in this work, provided as commented notebook files. Guidance on how to use these codes, interpret the animations, and explore the simulations further is provided in Sec.~\ref{supsec:codes_animations} of the Supplementary Notes.

With the computational framework now established, we turn to the physical phenomena revealed by these simulations. In the following sections, the solver is used to investigate wave scattering, frequency-selective transmission in layered media, and the emergence of band gaps in periodic structures.

\section{Wave Scattering at a Single Interface}
\label{sec:single_interface_scattering}

Band gaps in periodic media arise from repeated scattering and interference at material interfaces. To understand how such reflections arise, we begin with the simplest case: a single boundary between two homogeneous media. Although this problem is familiar from basic wave theory, examining it numerically serves an important pedagogical purpose. It establishes the computational treatment of interfaces and helps students become familiar with the numerical algorithm before moving on to periodic structures.

When a longitudinal elastic wave encounters a change in material properties, part of its energy is reflected while the remainder is transmitted. The quantitative description of this process is most conveniently expressed in terms of the mechanical impedance.

For elastic waves, the relevant fields are the stress $\sigma$ and the particle velocity $v$. The ratio of these quantities defines the mechanical impedance,\cite{kinsler2000}
\begin{equation}
Z = \frac{\sigma}{v}.
\end{equation}
For plane waves in a homogeneous medium, this reduces to
\begin{equation}
Z = \rho c,
\end{equation}
where $\rho$ is the mass density and $c$ is the wave speed. The impedance therefore characterizes the relation between stress and particle velocity in the medium.

The origin of reflection and transmission lies in the boundary conditions imposed on these fields. At a material interface, both particle velocity and stress must remain continuous. These constraints determine the reflected and transmitted amplitudes. When the impedances of the adjoining media differ ($Z_1 \neq Z_2$), the continuity conditions cannot be satisfied by a purely transmitted wave, and partial reflection necessarily occurs.

For normal incidence, the corresponding energy reflection and transmission coefficients are\cite{kinsler2000,yeh2005}
\begin{equation}
R = \left( \frac{Z_2 - Z_1}{Z_2 + Z_1} \right)^2,
\qquad
T = \frac{4 Z_1 Z_2}{(Z_1 + Z_2)^2},
\label{eq:energy_r_t}
\end{equation}
where $Z_1 = \rho_1 c_1$ and $Z_2 = \rho_2 c_2$ are the impedances of media 1 and 2. These coefficients depend only on the impedance contrast, highlighting that reflection is governed by material mismatch rather than the absolute values of the wave speed or density.

\begin{figure*}[tb]
\centering
\includegraphics[width=\linewidth]{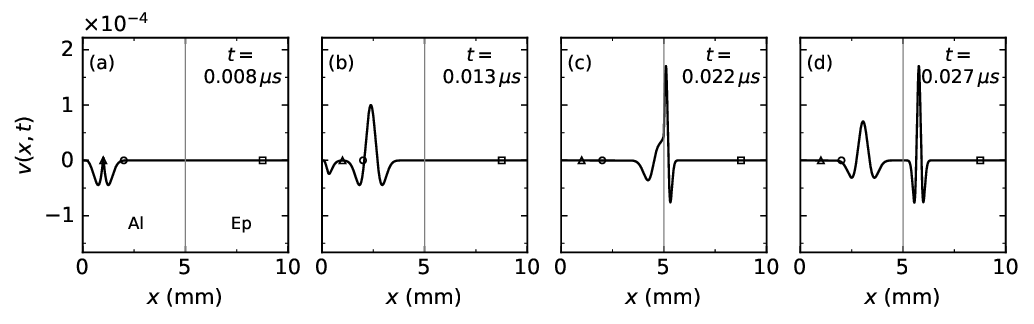}
\caption{Snapshots of the velocity field $v(x,t)$ as a pulse propagates from aluminium (Al) into epoxy (Ep). The vertical line marks the interface. The triangular marker indicates the source location; the circle (probe 1) and square (probe 2) denote the probe positions where the incident/reflected and transmitted signals are recorded. The sequence illustrates partial reflection in Al and transmission into Ep. Reflection and transmission coefficients extracted from the probe signals agree with the analytical impedance predictions.}
\label{fig:impedance_snaps}
\end{figure*}

These analytical expressions can be verified directly using the velocity--stress staggered-grid formulation developed earlier. Material parameters used in the simulations are listed in Table~\ref{tab:material_params}.\cite{aly2018,khateib2020} Figure~\ref{fig:impedance_snaps} shows snapshots of a pulse propagating from aluminium (Al) into epoxy (Ep). The vertical line marks the interface. The triangular marker indicates the source location, while the circle (probe 1) and square (probe 2) denote the probe positions where the incident/reflected and transmitted signals are recorded.

\begin{table}[ht]
\caption{\label{tab:material_params}
Material parameters used in the simulations.}
\begin{ruledtabular}
\begin{tabular}{lcc}
Material & Density $\rho$ ($\mathrm{kg\,m^{-3}}$) & Wave speed $c$ ($\mathrm{m\,s^{-1}}$) \\
\hline
Aluminium (Al) & 2700 & 6400 \\
Epoxy (Ep) & 1180 & 2535 \\
Water (W) & 1000 & 1490 \\
Lead (Pb) & 10760 & 1960 \\
\end{tabular}
\end{ruledtabular}
\end{table}

The reflected and transmitted energies are obtained from the instantaneous energy flux,
\begin{equation}
J(t) = \sigma(x,t)\, v(x,t),
\end{equation}
evaluated at the probe locations and integrated over the duration of the pulse. The reflectance and transmittance are defined as the ratios of reflected and transmitted energy to the incident energy. This formulation provides a direct check of energy conservation in the numerical scheme. Computational details are provided in Sec.~\ref{supsec:interface_energy_calc} of the Supplementary Notes.\cite{SuppMat}

Table~\ref{tab:impedance_results} compares the analytical and numerical values. The close agreement confirms that the interface conditions are correctly captured by the numerical scheme and that the solver preserves energy within the expected discretization error. This elementary scattering process forms the fundamental building block for wave propagation in layered media. The effect of repeated such reflections and transmissions will be examined in the next section.

\begin{table}[ht]
\caption{\label{tab:impedance_results}
Analytical and numerical energy reflection ($R$) and transmission ($T$) coefficients for a single Al--Ep interface.}
\begin{ruledtabular}
\begin{tabular}{lccc}
 & Analytical & Numerical & Error (\%) \\
\hline
$R$   & 0.49685 & 0.49672 & 0.026 \\
$T$   & 0.50315 & 0.50331 & 0.032 \\
\hline
$R+T$ & 1.00000 & 1.00003 & ---   \\
\end{tabular}
\end{ruledtabular}
\end{table}

\vspace{0.5em}
\noindent\textbf{Exercise 1.}
Repeat the interface-scattering simulation using a lead--epoxy material pair instead of aluminium--epoxy. Use the material parameters listed in Table~\ref{tab:material_params}. Compute the analytical reflection and transmission coefficients using Eq.~\eqref{eq:energy_r_t} and compare them with the numerical values obtained from the simulation. Examine how the larger impedance contrast modifies the partition of energy between reflected and transmitted waves.

\section{From Repeated Scattering to Frequency-Selective Transmission}
\label{sec:scattering_to_stopband}

\begin{figure}
\centering
\includegraphics[width=\linewidth]{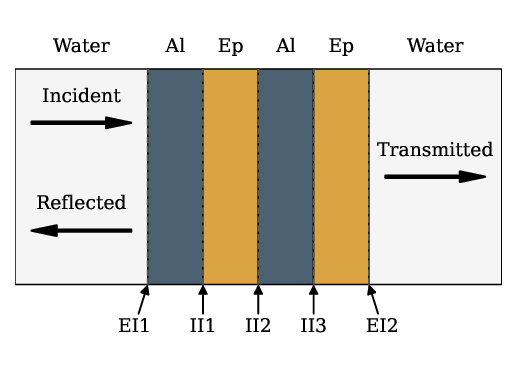}
\caption{Two-unit-cell layered structure embedded in water. EI1 and EI2 denote the external interfaces between water and the layered region, while II1--II3 denote internal Al--Ep interfaces. A wave incident from the left undergoes multiple partial reflections and transmissions at each interface.}
\label{fig:two_interface_schematic}
\end{figure}

Having established the scattering properties of a single impedance interface, we now examine how repeated interfaces reorganize wave transmission. To illustrate this process, we consider a short layered structure consisting of two unit cells (four layers) of alternating aluminium (Al) and epoxy (Ep), embedded in water. This minimal structure already contains multiple interfaces and therefore captures the essential mechanism of repeated scattering. The boundaries between water and the layered region are denoted EI1 and EI2 (external interfaces), while the Al--Ep interfaces inside the stack are labeled II1--II3 (internal interfaces). A schematic is shown in Fig.~\ref{fig:two_interface_schematic}.

When a pulse is incident from water onto EI1, partial reflection and transmission occur as discussed previously. The transmitted portion enters the layered region and encounters II1, where it again splits into reflected and transmitted components. These components propagate in both directions and subsequently interact with the remaining internal interfaces. Even for this short stack, multiple forward- and backward-propagating waves coexist within the structure.

Each traversal of a layer of thickness $d_i$ contributes a phase shift $k_i d_i$, where $k_i=\omega/c_i$ is the wavenumber in that material. The transmitted field observed beyond EI2 is therefore the coherent superposition of multiple contributions that have accumulated different propagation phases depending on how many internal reflections they have experienced. The transmission spectrum is governed by this phase-sensitive interference.\cite{yeh2005}

\begin{figure}[b]
\centering
\includegraphics[width=\linewidth]{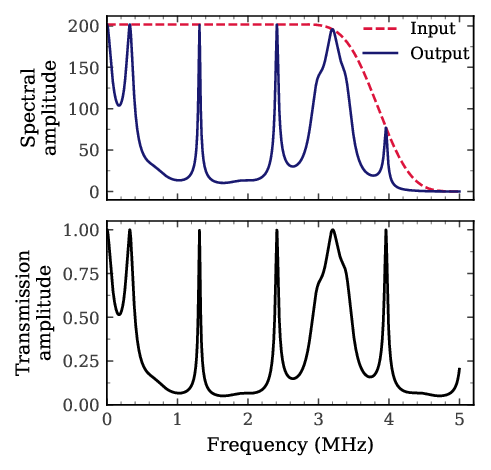}
\caption{Top: Input (dashed) and transmitted (solid) spectra for the two-unit-cell structure. Bottom: Normalized transmission. Multiple internal reflections reorganize broadband excitation into frequency-selective transmission.}
\label{fig:transmission_two_interface}
\end{figure}

Figure~\ref{fig:transmission_two_interface} shows the input spectrum, output spectrum, and normalized transmission for the two-unit-cell structure. Even with only four layers, the transmission is strongly frequency dependent. Sharp maxima and deep minima appear, demonstrating that repeated impedance mismatch reorganizes broadband excitation into a structured spectral response.\cite{sigalas1992}

Transmission maxima correspond to frequencies for which forward-propagating components inside the layered region arrive at EI2 with nearly identical phase, leading to constructive interference. Deep minima occur when internally reflected waves combine coherently in the backward direction, redirecting energy toward EI1 and suppressing net transmission.

These interference effects arise from two complementary mechanisms. Interference among reflections at the internal interfaces (II1--II3) results from waves confined within the layered region, while external-interface round trips involve waves that return to EI1 or EI2 and re-enter the stack after additional partial reflection. The latter mechanism produces narrow resonant features associated with the finite length of the stack, analogous to Fabry--Pérot resonances,\cite{born1999} whereas the broader transmission minima originate primarily from coherent phase relationships among internal reflections. As additional unit cells are introduced, the spectrum becomes increasingly structured: resonant features multiply and certain frequency intervals exhibit pronounced suppression.

The essential point is that repeated scattering reorganizes wave transport through phase correlations. Rather than simply attenuating the wave, the structure selectively enhances backward propagation over forward transmission in specific frequency ranges.

To make these phase relationships more explicit, consider the quarter-wavelength condition\cite{yeh2005} at a chosen design frequency $f_0$,
\begin{equation}
d_i=\frac{\lambda_i}{4}=\frac{c_i}{4f_0}.
\end{equation}

Under this condition, a round trip across any layer produces a phase shift
\begin{equation}
2k_i d_i=\pi.
\end{equation}

The consequence becomes clear by following the phase evolution of reflected components. Consider a wave incident from aluminium onto epoxy at II1. The amplitude reflection coefficient
\begin{equation}
r=\frac{Z_2-Z_1}{Z_2+Z_1}
\end{equation}
is negative because $Z_2<Z_1$, corresponding to a phase shift of $\pi$. The transmitted portion propagates across the epoxy layer, reflects at II2, and returns. The round-trip propagation contributes an additional phase $\pi$, while the reflection at II2 does not introduce a phase reversal. The returning component therefore acquires the same phase as the wave reflected earlier at II1.

Successive internal reflections thus emerge phase aligned. Backscattered components reinforce one another coherently, while forward-propagating components fail to build constructively. The cumulative phase condition across one unit cell,
\begin{equation}
k_1 d_1 + k_2 d_2 = \pi,
\label{eq:unit_cell_phase}
\end{equation}
therefore identifies a frequency at which the structure favors reflection over transmission.

This condition is closely related to Bragg interference in periodic media.\cite{kushwaha1993} For a unit-cell period $a=d_1+d_2$, constructive interference of reflected waves occurs when $k a \approx \pi$. In the quarter-wavelength design, each layer contributes a phase of $\pi/2$, so that the condition in Eq.~\eqref{eq:unit_cell_phase} is satisfied across a unit cell. Reflections from successive interfaces therefore add coherently, producing strong reflection and suppressing transmission near the design frequency.

\begin{figure}
\centering
\includegraphics[width=\linewidth]{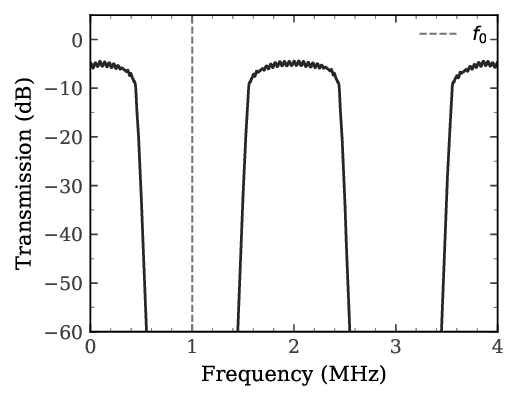}
\caption{Smoothed transmission (in dB) for a 15-unit-cell structure satisfying the quarter-wavelength condition at $f_0=1\,\mathrm{MHz}$. A deep minimum appears near $f_0$, where internal reflections become phase aligned.}
\label{fig:quarterwave_transmission}
\end{figure}

Figure~\ref{fig:quarterwave_transmission} shows the transmission of a 15-unit-cell structure designed so that each layer is one quarter wavelength at $f_0=1\,\mathrm{MHz}$.\cite{aly2018} A pronounced minimum occurs near this frequency, where the unit-cell phase condition is satisfied.

This frequency-selective suppression represents the finite-system precursor of a band gap in an infinite periodic medium. What begins as interference among a few interfaces acquires a collective character as additional layers are introduced. Even in a finite structure, periodic layering selects frequencies whose phase evolution suppresses sustained transmission. In the next section, this finite-stack phase condition is connected directly to the dispersion relation of an extended periodic medium.

\vspace{0.5em}
\noindent\textbf{Exercise 2.} 
Design a layered periodic structure that suppresses transmission near $2\,\mathrm{MHz}$ using the quarter-wavelength condition $d_i=c_i/(4f_0)$. Construct a multilayer stack satisfying this condition and compute the transmission spectrum. Verify that a pronounced transmission minimum occurs near the design frequency.

\vspace{0.5em}
\noindent\textbf{Exercise 3.}
Repeat Exercise 2 using a lead–epoxy bilayer instead of aluminium–epoxy while keeping the design frequency fixed at $2\,\mathrm{MHz}$. Compare the resulting band-gap width with that obtained for the aluminium–epoxy structure and examine how the increased impedance contrast influences the extent of the stop band.

\section{The Bloch Limit of Finite Periodic Structures}
\label{sec:bloch_limit}
Having examined wave transport through finite layered stacks, we now establish a direct correspondence with Bloch theory, which is traditionally derived under the assumption of infinite periodicity.\cite{ashcroft1976,kittel2004} In the following analysis we demonstrate how the phase relationships observed numerically in finite stacks provide a direct physical interpretation of the Bloch condition that governs perfectly periodic media.

The first conceptual point to clarify is the meaning of a stop band. In a finite stack, a deep transmission minimum does not imply that waves cease to exist at that frequency. Rather, the field penetrates the structure but attenuates rapidly with depth. The strong suppression observed numerically corresponds physically to spatial decay, not to the absence of fields.

In contrast, Bloch theory describes an infinite periodic medium. In such a system, solutions satisfy the Bloch condition
\begin{equation}
u(x+a)=u(x)e^{i k_B a},
\end{equation}
where $a=d_1+d_2$ is the unit-cell length, $u(x)$ denotes the wave field, and $k_B$ is the Bloch wavevector.\cite{ashcroft1976,kittel2004} When $k_B$ is real, waves propagate without attenuation and the frequency lies within a pass band. In a stop band, no real solution for $k_B$ exists. Instead,
\begin{equation}
k_B = k_B' + i k_B'',
\end{equation}
and the imaginary part $k_B''>0$ produces exponential decay,
\begin{equation}
u(x)\sim e^{-k_B'' x}.
\end{equation}
Thus, the infinite-medium statement that no propagating Bloch mode exists corresponds directly to the emergence of spatial attenuation.\cite{sigalas1992,kushwaha1993}

Bloch theory also predicts dispersion, which implies that the phase accumulated across one unit cell depends on frequency. In other words, the relation between the angular frequency $\omega$ (with $\omega = 2\pi f$) and the Bloch wavevector $k_B$ is not arbitrary but constrained by the periodic structure. For a bilayer system, this frequency dependence is given analytically by the Rytov relation for waves in periodic layered media,\cite{rytov1956,brekhovskikh1980,yeh2005}
\begin{multline}
\cos(k_B a)
=
\cos(k_1 d_1)\cos(k_2 d_2) \\
-
\frac{1}{2}
\left(
\frac{Z_2}{Z_1}
+
\frac{Z_1}{Z_2}
\right)
\sin(k_1 d_1)\sin(k_2 d_2)
\equiv D(\omega),
\label{eq:rytov_full}
\end{multline}
where $k_i=\omega/c_i$ and $Z_i=\rho_i c_i$ denote the wavenumber and acoustic impedance of each layer. Here $D(\omega)$ depends only on frequency and material parameters, and Eq.~\eqref{eq:rytov_full} reduces to
\begin{equation}
\label{eq:compact_rytov}
\cos(k_B a)=D(\omega).
\end{equation}
When $|D(\omega)|\le 1$, a real solution for $k_B$ exists and the frequency lies within a pass band. If instead $|D(\omega)|>1$, no real $k_B$ satisfies the equation. Taking the modulus of Eq.~\eqref{eq:compact_rytov} and writing $k_B=i k_B''$, with $\cos(i x)=\cosh(x)$, gives
\begin{equation}
\left|\cos(k_B a)\right|
=
\cosh(k_B'' a)
=
|D(\omega)|,
\end{equation}
so that
\begin{equation}
\label{eq:complex_k_bloch}
k_B''=
\frac{1}{a}
\cosh^{-1}\!\bigl(|D(\omega)|\bigr).
\end{equation}
The quantity $k_B''$ therefore specifies how rapidly the Bloch field decreases from one unit cell to the next when sustained phase propagation is not permitted.\cite{sigalas1992,kushwaha1993}

The Bloch relation in Eq.~\eqref{eq:compact_rytov} thus encodes two observable signatures of periodicity. In pass bands, where $|D(\omega)| \le 1$, Eq.~\eqref{eq:compact_rytov} admits real solutions for $k_B$, corresponding to propagating Bloch modes and a definite phase advance $k_B a$ across each unit cell. In stop bands, where $|D(\omega)|>1$, the Bloch wavevector becomes complex, yielding the attenuation constant $k_B''$ given in Eq.~\eqref{eq:complex_k_bloch}. We now examine both manifestations directly within the time-domain simulation and compare them with the analytical Rytov formulation.

First consider the attenuation signature. If our numerical development is consistent with Bloch theory, then for a frequency inside a numerically observed stop band, the spatial decay extracted from a finite stack should reproduce the analytical value of $k_B''$ within numerical accuracy. To test this, we excite a finite Al--Ep stack harmonically at $f_0=1\,\mathrm{MHz}$. The layer thicknesses were chosen according to the quarter-wavelength condition $d_i=c_i/(4f_0)$, so that $1\,\mathrm{MHz}$ lies within the first stop band. We compute the steady-state spatial amplitude by evaluating the root-mean-square (RMS) field over the stationary portion of the simulation. The envelope inside the periodic region is then fitted to
\begin{equation}
A(x)=A_0 e^{-k_{B,\mathrm{num}}'' x}.
\end{equation}
Details of the excitation design, steady-state windowing, RMS extraction, and fitting procedure are provided in Sec.~\ref{supsec:attenuation_dispersion} of the Supplementary Notes.\cite{SuppMat}

Figure~\ref{fig:evanescent_decay} shows the resulting spatial profile for a ten-unit-cell stack. The upper panel displays the oscillatory carrier, while the lower panel presents the same data on a logarithmic scale together with the exponential fit. The near-linear behavior on the logarithmic axis confirms that attenuation is approximately exponential over several unit cells.

\begin{figure}[t]
\centering
\includegraphics[width=\linewidth]{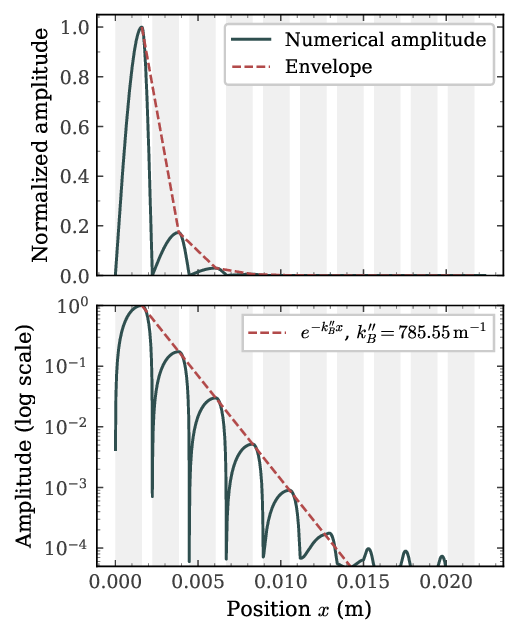}
\caption{Spatial amplitude inside a ten-unit-cell periodic stack for excitation at $f_0=1\,\mathrm{MHz}$ within the first stop band. The lower panel shows the exponential fit used to extract the numerical attenuation constant.}
\label{fig:evanescent_decay}
\end{figure}

For this case, the analytical prediction from Eq.~\eqref{eq:rytov_full} gives
\[
k_{B,\mathrm{Bloch}}'' = 785.16~\mathrm{m^{-1}},
\]
while the numerical fit yields
\[
k_{B,\mathrm{num}}'' = 785.55~\mathrm{m^{-1}},
\]
corresponding to a relative difference of approximately $0.05\%$. The close agreement confirms that the spatial attenuation observed in the finite stack is governed by the same Bloch attenuation constant predicted for an infinite periodic medium. In other words, the exponential decay measured directly in the time-domain simulation is not merely a numerical artifact of multiple reflections, but a manifestation of the evanescent Bloch mode associated with the stop band. This correspondence provides a direct bridge between the finite-stack scattering picture developed earlier and the Bloch-wave description of periodic media.

The attenuation signature confirmed the imaginary Bloch component. We now explore its complementary counterpart—the phase advance that organizes propagation across the lattice. From the Bloch condition,
\begin{equation}
\frac{u(x+a)}{u(x)}=e^{i k_B a},
\end{equation}
so that the phase advance across one unit cell directly yields $k_B(\omega)$ wherever propagation occurs. To extract this quantity numerically, we excite the structure using a broadband windowed-sinc pulse that spans the frequency range of interest. The resulting time-domain signals are recorded at two spatial locations separated by exactly one unit cell, and their cross-spectrum is computed. The complex argument (phase) of this cross-spectrum provides the phase difference between the two points, from which $k_B(\omega)$ is obtained. Computational details are summarized in Sec.~\ref{supsec:attenuation_dispersion} of the Supplementary Notes.\cite{SuppMat}

Figure~\ref{fig:rytov_numerical} compares the analytical Rytov dispersion with Bloch wavevectors extracted numerically for increasing total simulation time $T_f$. As $T_f$ increases, the frequency resolution improves and the numerical points populate the dispersion branches more densely across the pass bands. The emerging structure follows the analytical Rytov curves while clearly revealing the band gaps at the Brillouin-zone boundary ($ka=\pi$). This agreement shows that the phase advance measured in the time-domain simulation captures the dispersion relation predicted for an infinite periodic medium.

\begin{figure*}
\centering
\includegraphics[width=\textwidth]{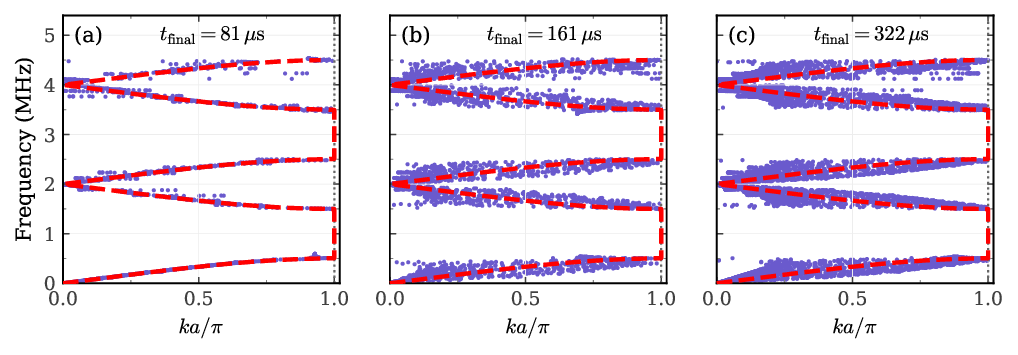}
\caption{Comparison between the analytical Rytov dispersion (dashed curves) and Bloch wavevectors extracted numerically (markers) for increasing total simulation time $T_f$. Longer simulation times improve frequency resolution, allowing the numerical points to populate the dispersion branches more densely and reveal the band gaps at the Brillouin-zone boundary.}
\label{fig:rytov_numerical}
\end{figure*}

The small deviations that remain are not unexpected. Our simulation uses a finite stack and a finite observation time. As discussed earlier, multiple reflections between the external interfaces introduce Fabry--Pérot–like modulations associated with the finite length of the structure.\cite{born1999} Finite time integration also limits the achievable frequency resolution. These effects are clearly visible near band edges and in strongly attenuated regions. They do not contradict Bloch theory; rather, they remind us that we are observing a numerical realization of a finite structure that approximates an ideal infinite periodic medium.

We therefore arrive at a coherent picture. What first appeared as interference among a few interfaces evolved into collective phase alignment across many layers. That collective behavior manifests, in the infinite limit, as a complex Bloch wavevector. The numerical algorithm bridges these descriptions directly: from repeated scattering, to exponential attenuation, to Bloch dispersion. The physics of stop bands thus emerges not as an abstract consequence of infinite periodicity, but as a structure already encoded within finite layered systems and revealed through careful phase-resolved computation.

\vspace{0.5em}

\noindent\textbf{Exercise 4.}
Repeat the attenuation analysis described above for the $2\,\mathrm{MHz}$ periodic structure designed in Exercise 3. Excite the structure harmonically near the stop-band frequency and extract the numerical decay constant $k_{B,\mathrm{num}}''$ by fitting the spatial envelope inside the periodic region to an exponential profile. Compare the value obtained with the attenuation predicted by the Bloch relation in Eq.~\eqref{eq:rytov_full}.

\vspace{0.5em}
\noindent\textbf{Exercise 5.}
Figure~\ref{fig:rytov_numerical} illustrates numerical artifacts in the extracted Bloch wave vector that arise from finite simulation conditions, particularly the choice of total simulation time $T_f$. Perform a complementary study by keeping the simulation time fixed while successively reducing the number of unit cells in the structure. Compute the numerical dispersion curves and compare the deviations from the analytical Rytov dispersion with those obtained when increasing the simulation time.

\section{Breaking Periodicity: Disorder and Defect States}
\label{sec:disorder_and_defect}

Throughout the discussion so far, we have considered perfectly periodic structures with fixed layer thicknesses and a repeating material sequence. Such ideal periodicity allowed us to understand how repeated scattering produces structured transmission and how this behaviour connects with Bloch theory. Real structures, however, rarely maintain perfect periodic order. Small variations in layer thickness may arise during fabrication, and structural perturbations may also be introduced intentionally to modify the spectral response. It is therefore instructive to examine how wave propagation responds when this periodic order is disturbed.

These situations can be explored directly within the same numerical framework by modifying the spatial material profile. Two representative perturbations are examined here: distributed random disorder across the lattice and a single localized defect embedded within an otherwise periodic structure.

We first examine distributed disorder. Independent random variations are introduced into the thickness of each layer of the quarter-wavelength phononic crystal described in Sec.~\ref{sec:scattering_to_stopband}. The structure consists of 15 unit cells. These fluctuations disturb the precise phase accumulation that normally occurs from one unit cell to the next. In a perfectly periodic stack, reflections from successive interfaces add coherently, producing well-defined pass bands and stop bands. When the layer widths vary randomly, the phase advance across each unit cell fluctuates, and the interference responsible for band formation gradually loses coherence.\cite{chen2007}

\begin{figure}[ht]
\centering
\includegraphics[width=\linewidth]{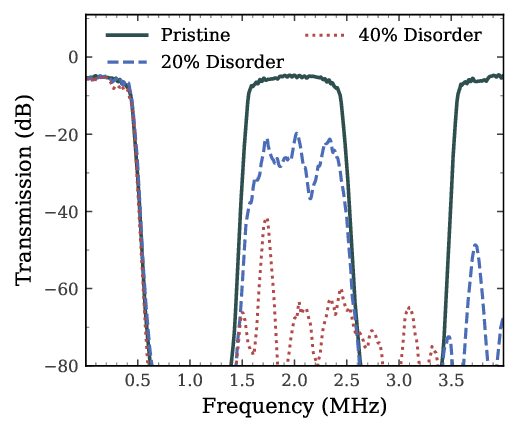}
\caption{Transmission spectrum of the quarter-wavelength phononic crystal under increasing random thickness disorder. Independent random variations of 20\% and 40\% are applied to the layer widths. Moderate disorder introduces fluctuations and reduces pass-band transmission, while stronger disorder progressively erodes the spectral structure by disrupting the phase coherence responsible for Bragg interference.}
\label{fig:spectral_erosion}
\end{figure}

Figure~\ref{fig:spectral_erosion} illustrates the resulting spectral evolution. For moderate disorder (20\%), the overall band structure remains recognizable, although pass-band transmission decreases and fluctuations appear. As the disorder increases to 40\%, the spectral features become fragmented and the band edges lose their sharpness. The principal stop band remains identifiable because the acoustic impedance contrast between layers is unchanged, but the coherent interference responsible for well-defined transmission bands is progressively degraded. Thus, increasing disorder weakens the phase coherence that underlies band formation.

While distributed disorder modifies the structure globally, a single localized perturbation produces a qualitatively different effect. Instead of introducing random variations in every unit cell, the modification is confined to one layer near the center of the crystal. In the present example, the central epoxy layer in an otherwise periodic six-cell stack is made twice as thick, creating a localized defect within the lattice.

A striking change appears in the transmission spectrum when such a defect is introduced. As discussed earlier, the surrounding periodic lattice strongly reflects waves within the stop-band frequency range, so a wave entering the structure normally decays rapidly rather than propagating through it. However, when the central layer is modified, the regular phase pattern of the crystal is locally disturbed. The doubled epoxy layer creates a region where the phase accumulated differs from that in the rest of the lattice. At a particular frequency, this region satisfies a standing-wave condition, allowing the wave to oscillate inside the defect while decaying into the surrounding periodic structure. The defect therefore supports a localized mode whose evanescent tails extend into neighboring layers.\cite{aly2018}

Because these decaying fields overlap with the incident and transmitted waves outside the crystal, energy can tunnel through the structure via this localized state, producing a narrow transmission peak inside the stop band. This process corresponds to resonant tunneling mediated by a localized defect mode.

\begin{figure}
\centering
\includegraphics[width=\linewidth]{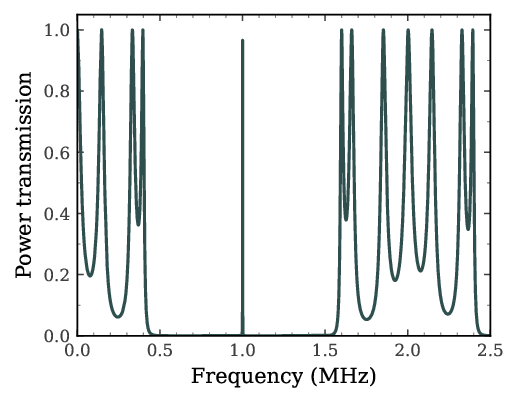}
\caption{Transmission spectrum of a six-cell quarter-wavelength phononic crystal with a single local defect created by doubling the thickness of the central epoxy layer. The surrounding periodic lattice blocks propagation across the stop band, while the defect introduces a localized resonant mode that appears as a narrow transmission peak inside the band gap.}
\label{fig:local_defect_state}
\end{figure}

Figure~\ref{fig:local_defect_state} shows the transmission spectrum for this configuration. The periodic lattice blocks propagation across the stop-band region, but the defect introduces a single transmission channel within that forbidden frequency range. The situation is closely analogous to impurity states in semiconductor band gaps, where a localized perturbation creates an allowed state inside the gap. Here, the same physics appears for elastic waves: the periodic structure forms the band gap, and the defect locally creates a resonant mode within it.

These examples illustrate how strongly wave transport in periodic media depends on structural order. Distributed disorder gradually weakens the phase coherence responsible for band formation, while a localized defect introduces new allowed states inside an otherwise forbidden spectral region. Both effects arise from the same interference physics that governs propagation in periodic media, highlighting how scattering and phase accumulation together control the spectral response. In this way, disorder and defects provide a natural extension of Bloch theory, illustrating how deviations from perfect periodicity modify wave transport through the same underlying interference mechanisms.

\vspace{0.5em}

\noindent\textbf{Exercise 6.}
Extend the disorder study shown in Fig.~\ref{fig:spectral_erosion} by introducing random variations in the material properties in addition to the layer thicknesses. First apply small random fluctuations to $\rho$ and $c$, and then combine these with thickness disorder. Compute the resulting transmission spectra and examine how variations in material properties influence the degradation of the band structure.

\vspace{0.5em}
\noindent\textbf{Exercise 7.}
Repeat the defect study by modifying the defect thickness using other scaling factors (e.g., $1.5$ and $2.5$ times the original layer width). Compute the transmission spectra and observe how the defect-mode frequency shifts as the defect geometry changes.

\section{Conclusion}

We have presented a compact computational framework through which band formation in periodic media can be reconstructed directly from time-domain wave dynamics in finite layered systems. Beginning with scattering at a single interface and progressing to multilayer interference, the simulations show how repeated reflections and phase accumulation organize wave propagation and give rise to the attenuation and transmission patterns associated with band structures. In this way, the Bloch description of an infinite periodic medium emerges naturally from the dynamics of finite stacks, providing a clear connection between familiar time-domain wave processes and the more abstract band concepts typically introduced through dispersion relations. In this sense, Bloch theory is not an independent starting point, but a natural consequence of wave interference in periodic media.

Pedagogically, the framework serves as a computational laboratory for teaching wave propagation in periodic media. The minimal solver allows students to vary structural parameters and directly observe their influence on transmission, attenuation, and band formation. Through such investigations, students gain experience with numerical modeling while developing physical intuition through visualization of wave dynamics. The framework is simple yet transferable, providing a practical tool that can be extended to a wide range of wave phenomena and more advanced studies of structured media.

\section{Extensions and Further Exploration}
\label{sec:extensions}

The computational framework developed here enables several natural extensions and further explorations, a few of which are outlined below.

\begin{enumerate}

\item The present solver readily supports a variety of additional one-dimensional studies. Once implemented, students can modify layer thicknesses, material properties, or introduce defects to investigate band-gap tuning, localized defect modes, and the influence of impedance contrast on transmission in periodic layered media.\cite{elboudouti2018,kushwaha1993,chaudhary2026,wu2017,lazcano2014}

\item The staggered-grid update scheme follows the leapfrog finite-difference time-domain (FDTD) method originally introduced by Yee for electromagnetic waves. In the present context, it is applied to the velocity and stress fields of elastodynamics. This connection provides a natural pathway to extend the framework to electromagnetic wave simulations; a clear pedagogical treatment is given in Ref.~\onlinecite{sipos2008}.

\item The one-dimensional model can also be extended to two dimensions. Such simulations enable exploration of richer phononic or photonic crystal phenomena, including directional band gaps, waveguiding, and defect cavities.\cite{pennec2018,norris,lou2004} Since phononic crystals are an active area of research with applications in vibration control and wave manipulation, the framework developed here can also serve as a starting point for research-oriented undergraduate projects.

\end{enumerate}

\section*{Authors Declaration}

The authors declare that they have no conflicts of interest.

\end{document}